\documentclass[pdftex,twocolumn,epjc3]{svjour3}          

\RequirePackage[T1]{fontenc}
\usepackage{lineno}

\smartqed  

\RequirePackage{graphicx}
\RequirePackage{mathptmx}      
\RequirePackage{flushend}
\RequirePackage[numbers,sort&compress]{natbib}
\RequirePackage[colorlinks,citecolor=blue,urlcolor=blue,linkcolor=blue]{hyperref}

\begin{document}


\title{Measuring the leading hadronic contribution to the muon $g$-2 via $\mu e$ scattering}

\author{G.~Abbiendi\thanksref{e1,addr1}
        \and
        C.~M.~Carloni~Calame\thanksref{e2,addr2} 
        \and
        U.~Marconi\thanksref{e3,addr3} 
        \and
        C.~Matteuzzi\thanksref{e4,addr4} 
        \and
        G.~Montagna\thanksref{e5,addr5,addr2} 
        \and
        O.~Nicrosini\thanksref{e6,addr2} 
        \and
        M.~Passera\thanksref{e7,addr6} 
        \and
        F.~Piccinini\thanksref{e8,addr2} 
        \and
        R.~Tenchini\thanksref{e9,addr7} 
        \and
        L.~Trentadue\thanksref{e10,addr8, addr4} 
        \and
        G.~Venanzoni\thanksref{e11,addr9} 
}

\thankstext{e1}{e-mail: giovanni.abbiendi@bo.infn.it}
\thankstext{e2}{e-mail: carlo.carloni.calame@pv.infn.it}
\thankstext{e3}{e-mail: umberto.marconi@bo.infn.it}
\thankstext{e4}{e-mail: clara.matteuzzi@cern.ch}
\thankstext{e5}{e-mail: guido.montagna@pv.infn.it}
\thankstext{e6}{e-mail: oreste.nicrosini@pv.infn.it}
\thankstext{e7}{e-mail: massimo.passera@pd.infn.it}
\thankstext{e8}{e-mail: fulvio.piccinini@pv.infn.it}
\thankstext{e9}{e-mail: roberto.tenchini@cern.ch}
\thankstext{e10}{e-mail: luca.trentadue@cern.ch}
\thankstext{e11}{e-mail: graziano.venanzoni@lnf.infn.it}

\institute{INFN Bologna, Viale Carlo Berti-Pichat 6/2, 40127 Bologna, Italy \label{addr1}
          \and
          INFN Pavia, Via Agostino Bassi 6, 27100 Pavia, Italy \label{addr2}
          \and
          INFN Bologna, Via Irnerio 46, 40126 Bologna, Italy \label{addr3}
          \and
          INFN Milano Bicocca, Piazza della Scienza 3, 20126 Milano, Italy \label{addr4}
          \and
          Dipartimento di Fisica, Universit\`a di Pavia, Via A. Bassi 6, 27100, Pavia, Italy \label{addr5}
          \and
          INFN Padova, Via Francesco Marzolo 8, 35131 Padova, Italy \label{addr6}
          \and
          INFN Pisa, Largo Bruno Pontecorvo 3, 56127 Pisa, Italy \label{addr7}
          \and
          Dipartimento di Fisica e Scienze della Terra ``M. Melloni", Parco Area delle Scienze 7/A, 43124 Parma, Italy \label{addr8}
          \and
          INFN, Laboratori Nazionali di Frascati, Via E. Fermi 40, 00044 Frascati (RM), Italy \label{addr9}
}


\maketitle

\begin{abstract}
We propose a new experiment to measure the running of the electromagnetic coupling constant  
in the space-like region by scattering high-energy muons on atomic electrons of a low-{\it Z} target through
the elastic process $\mu \, e \to \mu \, e$. The differential cross section of this
process, measured as a function of the squared momentum transfer \mbox{$t=q^2<0$}, 
provides direct sensitivity to the leading-order hadronic contribution to the muon anomaly $a^{\rm{HLO}}_{\mu}$. 
By using a  muon beam of 150~GeV, with an
average rate of $\sim1.3\times 10^7$ muon/s, currently available at the
CERN North Area, a statistical uncertainty of {$\sim 0.3\%$} can be
achieved on $a^{\rm{HLO}}_{\mu}$ after two years of data taking. 
The direct measurement of $a^{\rm{HLO}}_{\mu}$ via $\mu e$ scattering will provide an independent
determination, competitive with the time-like dispersive approach, and consolidate the theoretical 
prediction for the muon $g$-2 in the Standard Model. It will allow therefore a firmer 
interpretation  of the measurements of the future muon $g$-2 
experiments at Fermilab and J-PARC.
\end{abstract}
\section{Introduction}
In searching for new physics, low-energy high-precision measurements are complementary to the LHC high-energy frontier. 
The long-standing (3--4)$\sigma$ discrepancy between the experimental value of the muon anomalous magnetic moment $a_{\mu}=(g-2)/2$ 
and the Standard Model (SM) prediction, \mbox{$\Delta a_{\mu}(\rm Exp-SM) \sim (28\pm 8)\times 10^{-10}$}~\cite{Blum:2013xva,Jegerlehner:2015stw}, 
has been considered during these years as one of the most intriguing indications of physics beyond the SM.
However, the accuracy of the SM prediction, $5\times 10^{-10}$, is limited by strong interaction effects, 
which cannot be computed perturbatively at low energies. 
Long time ago, by using analyticity and unitarity, it was shown \cite{BM61GDR69} that the leading-order (LO) hadronic contribution to the muon $g$-2, $a_{\mu}^{\rm{HLO}}$, 
could be computed via a dispersion integral of the hadron production cross section in $e^+e^-$ annihilation at low-energy. 
The present error on $a_{\mu}^{\rm{HLO}}$, $\sim 4\times 10^{-10}$, with a fractional accuracy of 0.6\%, constitutes the main uncertainty of the SM prediction. 
An alternative evaluation of $a_{\mu}^{\rm{HLO}}$ can be obtained by lattice QCD calculations~\cite{lattice}. 
Even if current lattice QCD results are not yet competitive with those obtained with the dispersive approach via time-like data, 
their errors are expected to decrease significantly in the next few years~\cite{Blum:2013qu}.
The ${\cal O}(\alpha^3)$ hadronic light-by-light contribution, $a_{\mu}^{\rm{HLbL}}$, 
which has the second largest error in the theoretical evaluation, 
contributing with an uncertainty of (2.5--4)$\times 10^{-10}$, cannot at present be determined from data and its calculation relies on the use of specific models~\cite{Jegerlehner:2009ry, HLBL}. 

From the experimental side, the error achieved by the BNL E821 experiment, $\delta a_{\mu}^{\rm{Exp}}= 6.3 \times 10^{-10}$ (corresponding to 0.54 ppm)~\cite{Bennett:2006fi}, is dominated by the available statistics. New experiments
at Fermilab and J-PARC, aiming at measuring the muon $g$-2 to a precision of $1.6 \times 10^{-10}$ (0.14 ppm), are in preparation~\cite{Grange:2015fou,Saito:2012zz}.
%
%
\begin{figure}[t]
\begin{center}
  \includegraphics[width=0.5\textwidth,angle=0]{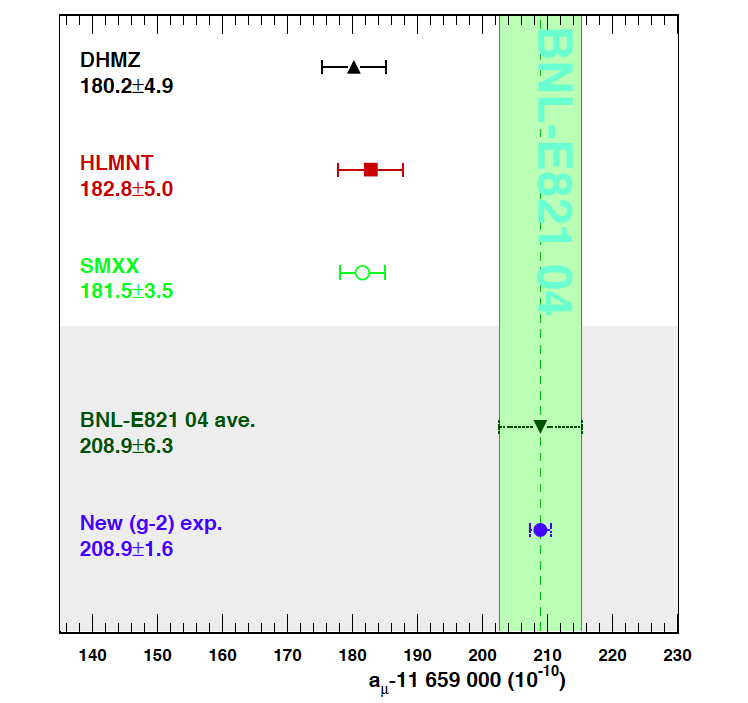}
 \caption{Comparison between the SM predictions and the experimental determinations $a_{\mu}^{\rm SM}$ and $a_{\mu}^{\rm Exp}$. 
DHMZ is Ref.~\cite{Davier11}, HLMNT is Ref.~\cite{Hagiwara:2011}; 
SMXX  \cite{SMXX} is the average of the two previous values with a reduced error as expected by the improvement on the hadronic cross section measurement; BNL-E821 04 ave. is the current 
experimental value of $a_{\mu}$;  New ($g$-2) exp. is the same central value with a fourfold improved precision, as planned by the future $g$-2 experiments at Fermilab and J-PARC \cite{Blum:2013xva}.}
\label{fg:g-2new-v3}
\end{center}
\end{figure}
Fig.~\ref{fg:g-2new-v3}, from Ref.~\cite{Blum:2013xva}, shows the status of the $g$-2 discrepancy compared with what could be expected after the new $g$-2 measurements at Fermilab and J-PARC, assuming that the central value would remain the same. 
Together with a fourfold improved precision on the experimental side, an improvement on the LO hadronic contribution is highly desirable. 
Differently from the dispersive approach, which relies on time-like data from annihilation cross sections, 
our proposal is to determine $a_{\mu}^{\rm{HLO}}$ from a measurement of the effective electromagnetic coupling in the space-like region, 
where the vacuum polarization is a smooth function of the squared momentum transfer.  
This method has been recently proposed~\cite{Calame:2015fva} by using Bhabha scattering data. 
A method to determine the running of $\alpha$ by using small-angle Bhabha scattering was proposed in~\cite{Arbuzov:2004wp} and applied to LEP data in~\cite{Abbiendi:2005rx}. 
The hadronic contribution to the running of $\alpha$ can also be determined unambiguously through the $t$-channel $\mu e$ elastic scattering process, 
from which $a_{\mu}^{\rm{HLO}}$ could be obtained, as detailed in this paper.

The paper is organized as follows. After a short review of the theoretical framework in Sect.~\ref{sec:theory}, we present our experimental proposal 
in Sect.~\ref{sec:exp}. Preliminary considerations on the detector and systematic uncertainties are given in Sect.~\ref{sec:det} and Sect.~\ref{sec:unc}, 
respectively, while our conclusions are drawn in Sect.~\ref{sec:concl}.
\section{Theoretical framework}
\label{sec:theory}

With the help of dispersion relations and the optical theorem, the LO hadronic contribution to the muon $g$-2 is given by the well-known formula~\cite{BM61GDR69,Jegerlehner:2008zza}
\begin{equation}\label{amu}
        a_{\mu}^{\rm HLO} = \left( \frac{\alpha m_\mu}{3\pi}\right)^2\int_{4m_\pi^2}^{\infty} ds
        \;\frac{\hat{K}(s) R_{\rm had}(s)}{s^2},
\end{equation}
where $R_{\rm had}(s)$ is the ratio of the total $e^+e^-\to {\rm hadrons}$ and
the Born $e^+e^-\to\mu^+\mu^-$ cross sections, $\hat{K}(s)$ is a smooth function and $m_\mu$ ($m_\pi$) is the muon (pion) mass.
We remark that $R_{\rm had}(s)$ in the integrand function of Eq.~(\ref{amu}) is highly fluctuating at low energy due to hadronic resonances and threshold effects.
The dispersive integral in Eq.~(\ref{amu}) is usually calculated by using the experimental value of $R_{\rm had}(s)$ up to a certain value of $s$~\cite{Jegerlehner:2009ry, Davier:2010nc, Hagiwara:2011af} and by using perturbative QCD (pQCD)~\cite{pQCD} in the high-energy tail.
For the calculation of $a_{\mu}^{\rm HLO}$, an alternative formula can also be exploited~\cite{Lautrup:1971jf,Calame:2015fva}, namely
\begin{equation}\label{amu_xalpha}
        a_{\mu}^{\rm HLO} = 
         \frac{\alpha}{\pi} \int_0^1 dx \, (1-x) \,  \Delta \alpha_{\rm had} \! \left[ t(x) \right],
\end{equation}
where $\Delta\alpha_{\rm had}(t)$ is the hadronic contribution to the running of the fine-structure constant, evaluated at 
\begin{equation}
        t(x)=\frac{x^2m_\mu^2}{x-1} < 0,
\label{t}
\end{equation}
the space-like (negative) squared four-momentum transfer.
In contrast with the integrand function of Eq. (\ref{amu}), the integrand in Eq. (\ref{amu_xalpha}) is smooth and free of resonances.

By measuring the running of $\alpha$,
\begin{equation}\label{eq:alphaq2}
        \alpha(t) = \frac{\alpha(0)}{1-\Delta \alpha(t)},
\end{equation}
where $t=q^2<0$ and $\alpha(0)=\alpha$ is the fine-structure constant in the Thomson limit,
the hadronic contribution $\Delta\alpha_{\rm had}(t)$ can be extracted by subtracting from $\Delta \alpha(t)$ the purely leptonic part $\Delta \alpha_{\rm lep} (t)$,
which can be calculated order-by-order in perturbation theory (it is known up to three loops in QED \cite{Steinhauser:1998rq} and up to four loops in specific $q^2$ limits~\cite{4loop}).

\begin{figure*}[htp]
\includegraphics[width=.5\textwidth]{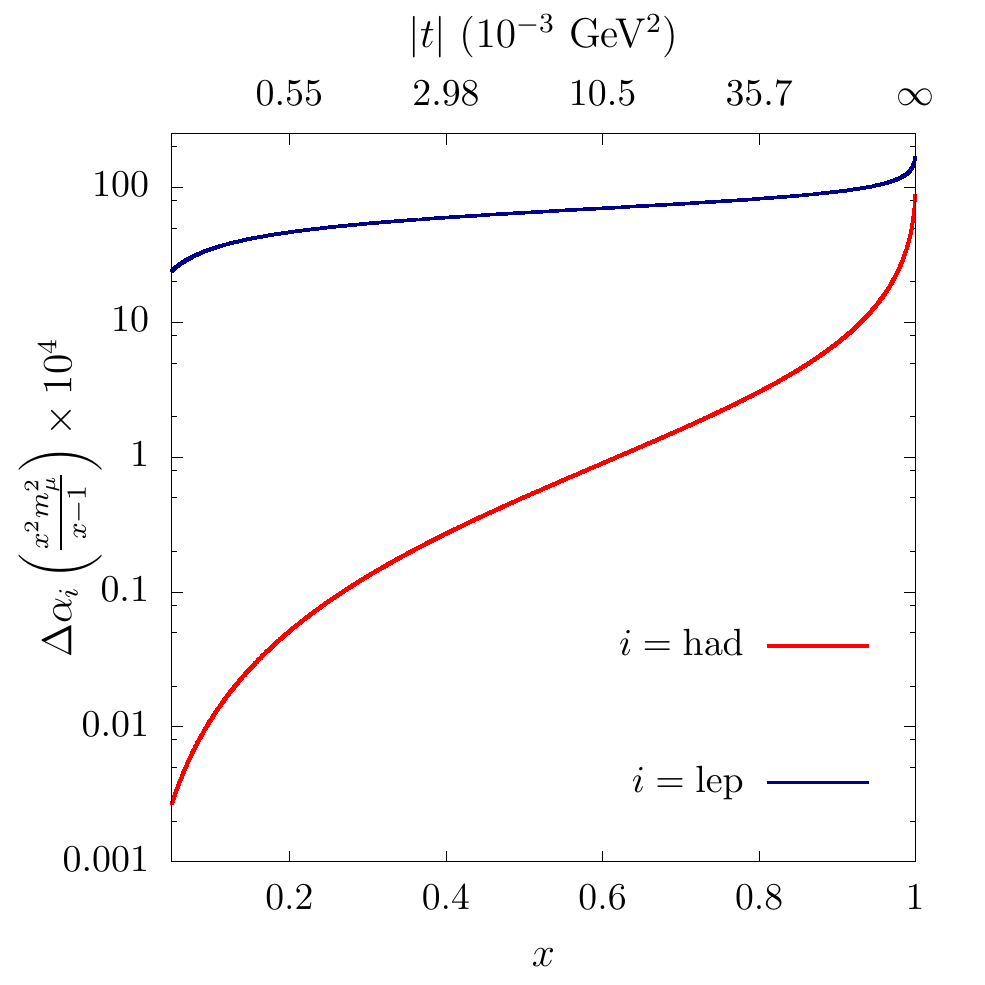}~\includegraphics[width=.5\textwidth]{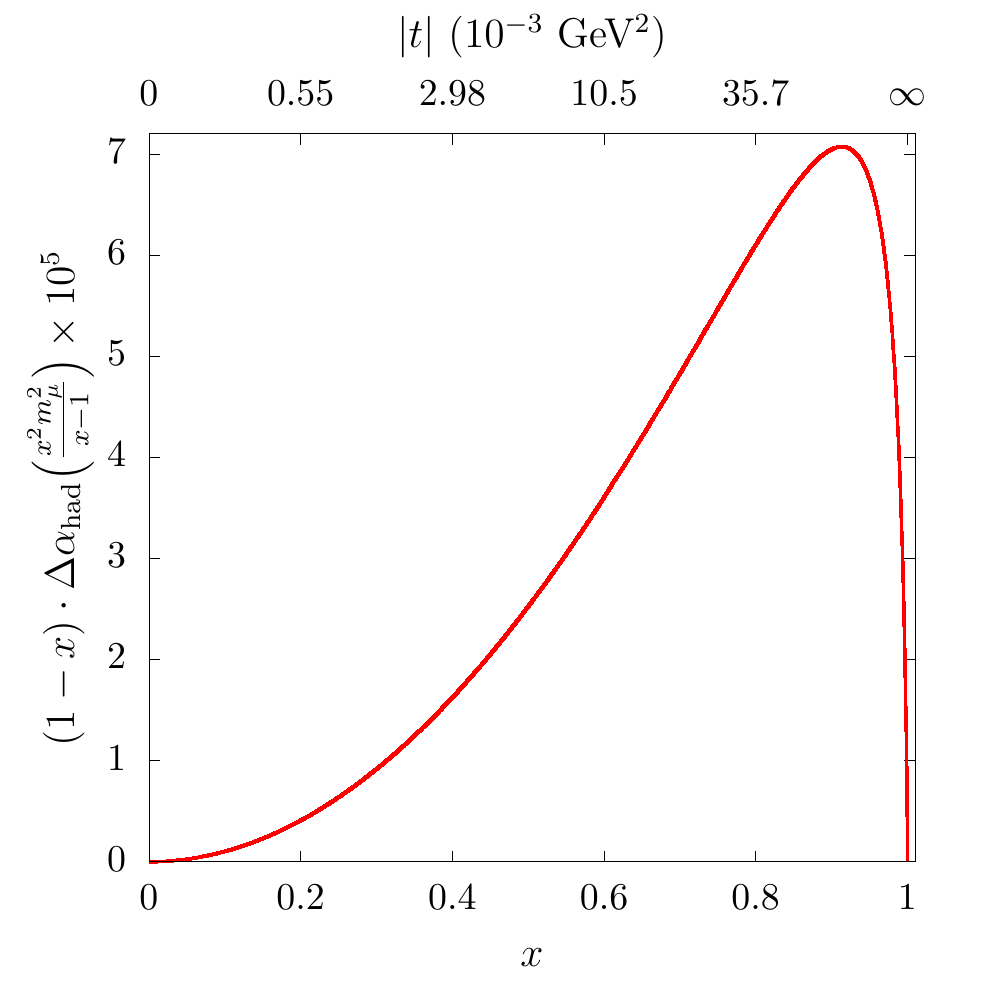}
  \caption{
    Left: $\Delta\alpha_{\rm had}[t(x)] \times 10^4$ (red) and, for comparison, $\Delta\alpha_{\rm lep}[t(x)] \times 10^4$ (blue), 
    as a function of $x$ and $t$ (upper scale).
    Right: the integrand $(1-x)\Delta\alpha_{\rm had}[t(x)] \times 10^5$
    as a function of $x$ and $t$. The peak value is at 
    $x_{\rm peak}\simeq 0.914$, corresponding to $t_{\rm peak} \simeq-0.108$ GeV$^2$.
}
  \label{xfunction}
\end{figure*}

Fig.~\ref{xfunction} (left) shows $\Delta\alpha_{\rm lep}$ and $\Delta\alpha_{\rm had}$ as a function of the variables $x$ and $t$. 
The range $x \in (0,1)$ corresponds to $t \in (-\infty,0)$, with $x=0$ for $t=0$.
The integrand of Eq.~(\ref{amu_xalpha}), calculated with the routine \texttt{hadr5n12}~\cite{dalphaFred}, 
which uses time-like hadroproduction data and perturbative QCD, is plotted in Fig.~\ref{xfunction} (right). 
The peak of the integrand occurs at $x_{\rm peak}\simeq 0.914$ (corresponding to $t_{\rm peak} \simeq-0.108$ GeV$^2$) 
and $\Delta\alpha_{\rm had}(t_{\rm peak}) \simeq 7.86 \times {10}^{-4}$ (see Fig.~\ref{xfunction} (right)).

\section{Experimental proposal}
\label{sec:exp}

We propose to use Eq.~(\ref{amu_xalpha}) to determine $a_{\mu}^{\rm{HLO}}$ by measuring the running of $\alpha$ in the space-like region  
with a muon beam of $E_{\mu}=150$~GeV on a fixed electron target.
The proposed technique is similar to the one used for the measurement of the pion form factor, as described in \cite{NA7ref}. 
It is very appealing for the following reasons:

$(i)$ It is a $t$-channel process, making the dependence on $t$ of the differential cross section proportional to $\left| \alpha(t) / \alpha(0) \right |^2$:
\begin{equation}
  \frac{d\sigma}{dt} = \frac{d\sigma_0}{dt}\left|\frac{\alpha(t)}{\alpha(0)}\right|^2,
\label{eq:sigmaLO}
\end{equation}
where $d\sigma_0/dt$ is the effective Born cross section, including virtual and soft photons, 
analogously to Ref.~\cite{Arbuzov:1997}, where small-angle Bhabha scattering at high energy was considered.
The vacuum polarization effect, in the leading photon $t$-channel exchange, is incorporated in the running of $\alpha$
and gives rise to the factor $\left| \alpha(t)/\alpha(0) \right|^2$.
It is understood that for a high precision measurement also higher-order radiative corrections must be included. 
For a detailed discussion see Refs.~\cite{Arbuzov:1997,Arbuzov:2004wp}.

$(ii)$ Given the incoming muon energy $E_\mu^i$, in a fixed-target experiment the $t$
variable is related to the energy of the scattered electron $E_e^f$ or its angle
 $\theta_e^f$:
\begin{equation}
t = (p_\mu^i-p_\mu^f)^2 = (p_e^i-p_e^f)^2 = 2m_e^2 - 2m_eE_e^f,
\end{equation}
\begin{equation}
s = (p_\mu^f+p_e^f)^2 = (p_\mu^i+p_e^i)^2 = m_\mu^2 + m_e^2 + 2m_eE_\mu^i,
\end{equation}
\begin{equation}
E_e^f = m_e\frac{1+r^2c^2_e}{1-r^2c^2_e}, \quad
\theta_e^f = \arccos\left(\frac{1}{r}\sqrt{\frac{E_e^f-m_e}{E_e^f+m_e}}\right),
\label{eq:kinematics}
\end{equation}
where
\begin{equation}
r\equiv \frac{\sqrt{(E^{i}_\mu)^2-m^2_\mu}}{E^i_\mu + m_e},
\quad c_e \equiv \cos\theta_e^f;
\label{eq:kinedefs}
\end{equation}

\noindent The angle $\theta_e^f$ spans the range (0--31.85)~mrad
for the electron energy $E_e^f$ in the range (1--139.8)~GeV
(the low-energy cut at 1 GeV is arbitrary).
%
%
\begin{figure*}[t]
\begin{center}
\includegraphics[width=1.0\textwidth,angle=0]{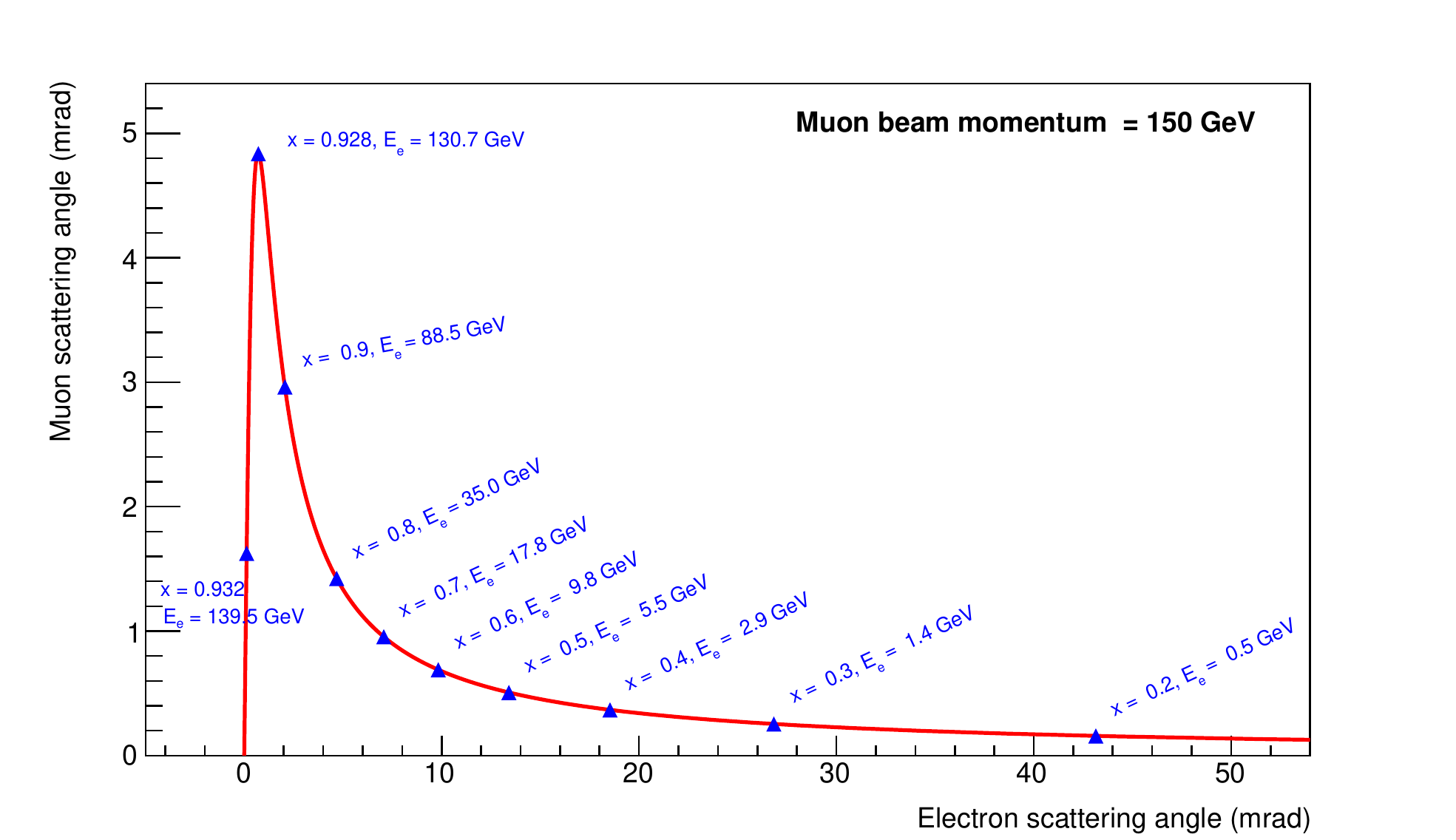}
 \caption{The relation between the muon and electron scattering angles for 150 GeV
 incident muon beam momentum. Blue triangles indicate reference values of the Feynman's $x$ and electron energy.}
\label{giovanni}
\end{center}
\end{figure*}
%

$(iii)$ For $E_\mu^i=150$~GeV, it turns out that $s\simeq 0.164$~GeV$^2$ and $-0.143$~GeV$^2 < t <0$~GeV$^2$
 (i.e. $-\lambda(s,m_\mu^2,m_e^2)/s <t < 0$, where $\lambda(x,y,z)$ is the
  K\"all\'en function). 
It implies that the region of $x$ extends up to 0.93, while the peak of the integrand function of Eq.~(\ref{amu_xalpha}) is at $x_{peak}=0.914$, 
corresponding to an electron scattering angle of 1.5 mrad, as visible in Fig.~\ref{xfunction} (right).

$(iv)$ The angles of the scattered electron and muon are correlated as
  shown in Fig.~\ref{giovanni} (drawn for incoming muon energy of $150$~GeV).
This constraint is extremely important to select elastic scattering
events, rejecting background events from radiative or inelastic
processes and to minimize systematic effects in the determination of
$t$. Note that for scattering angles of (2--3)~mrad there is 
an ambiguity between the outgoing electron and muon, as their angles and
momenta are similar, to be resolved by means of $\mu / e$ discrimination. 

$(v)$ The boosted kinematics allows the same detector to cover the whole acceptance. 
Many systematic errors, {\it e.g.} on the efficiency, will cancel out (at least at
first order) in the relative ratios of event counts in the high and
low $q^2$ regions (signal and normalization regions).

\begin{figure*}[htp]
\includegraphics[width=.5\textwidth]{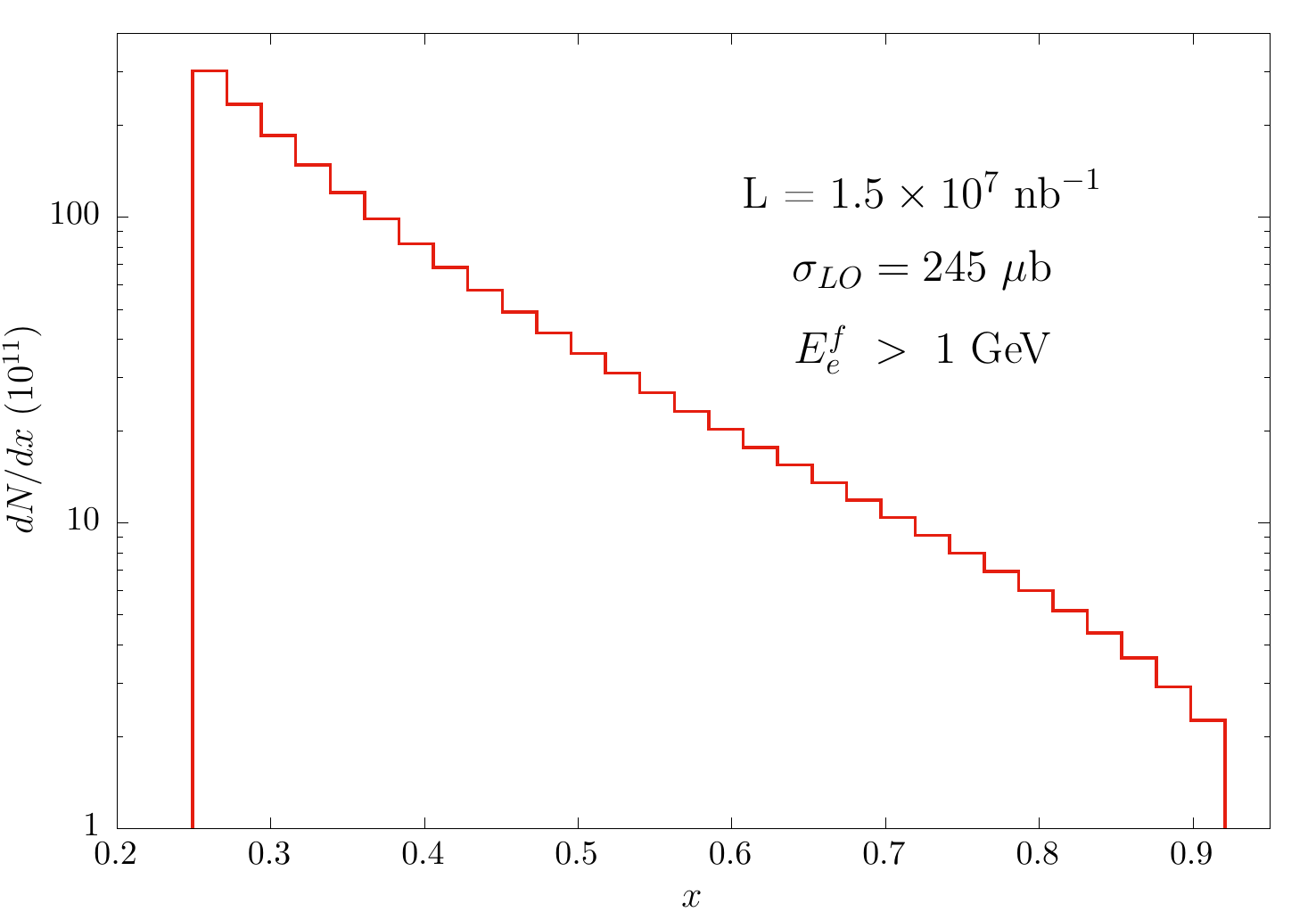}~\includegraphics[width=.5\textwidth]{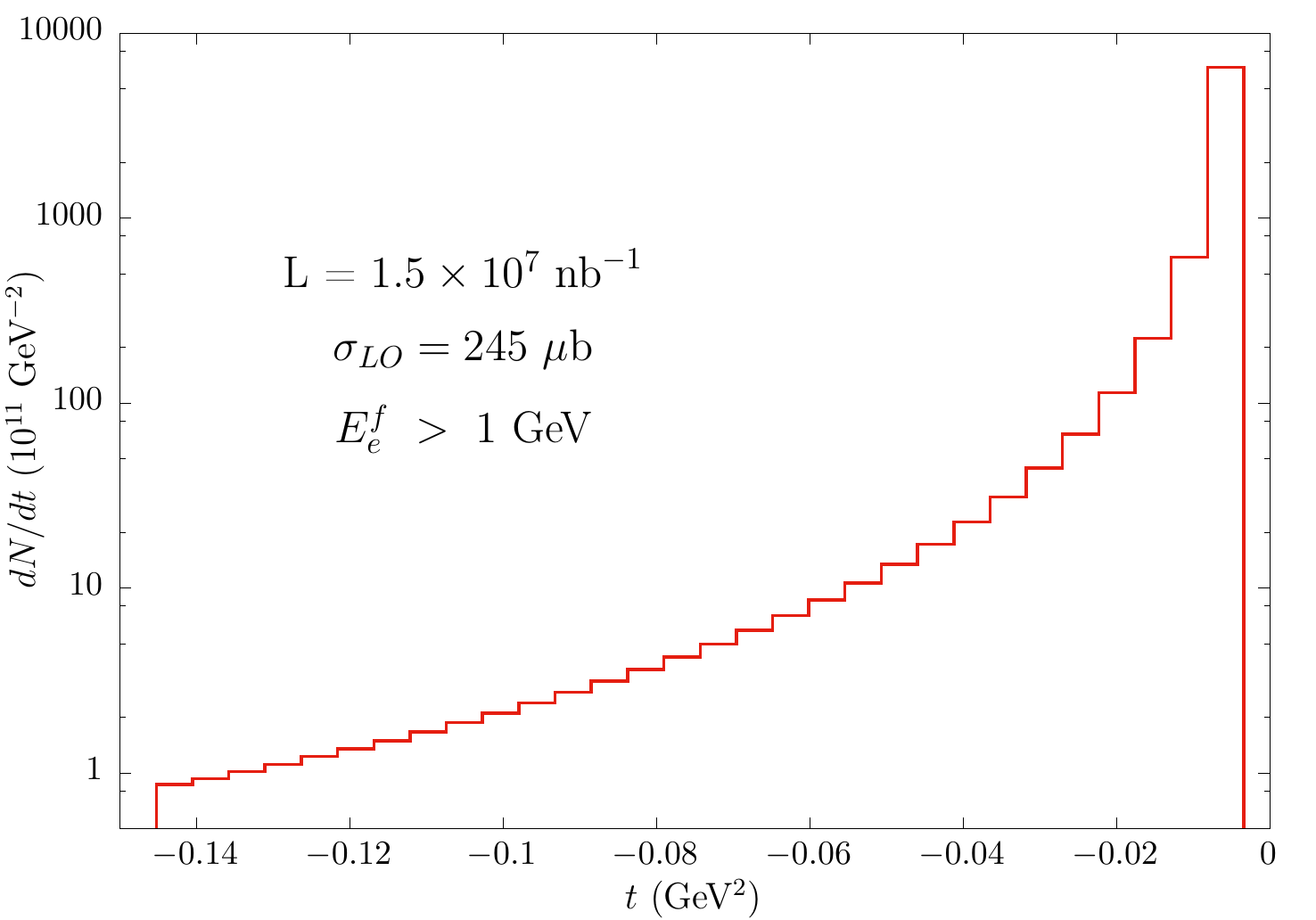}
\caption{
	Distribution of $\mu$-e scattering events as a function of $x$ (left) and $t$ (right) assuming 
	the integrated luminosity \mbox{$L = 1.5 \times 10^{7}$~nb$^{-1}$} and the LO elastic cross-section for $E_{e} > 1$~GeV, $\sigma_{LO} = 245$~$\mu$b.
}
  \label{dsdxdsdt}
\end{figure*}

Assuming a 150~GeV muon beam with an average intensity of $\sim 1.3 \times 10^7$ muons/s, presently available at CERN's North Area \cite{GatignonPrivate}, 
incident on a target consisting of twenty Beryllium layers, each 3~cm thick (see Sect.~4), and two years of data taking with a running time of $2 \times 10^7$~s/yr, 
one can reach an integrated luminosity of about $1.5 \times 10^7~{\rm nb}^{-1}$.
Taking into account the process cross section and the above value for the integrated luminosity, 
with a simplified simulation of the experiment we estimate that one can reach a statistical sensitivity of roughly $0.3\%$ on the value of $a_\mu^{\rm HLO}$. 
We considered 30 experimental data points in the accessible $x$ range. 
The integrand in the region $x \in [0.93, 1]$, accounting for $13\%$ of the $a_\mu^{\rm HLO}$ integral, cannot be reached by the proposed experiment, but  
can be determined using time-like data and perturbative QCD, and/or lattice QCD results \cite{RefJ,Marina1, Marina2}.
In Fig.~\ref{dsdxdsdt}, the distribution of the events, expected  with the above luminosity, is shown as a function of $x$ (left) and of $t$
(right), as obtained with a simulation of the lowest-order  $\mu e \to \mu e$ cross section in 30 evenly spaced bins. 
As can be seen, the statistics accessible in the $x$ peak region, corresponding to large momentum transfer $t$, is not a limiting factor.
\section{Preliminary considerations on the detector}
\label{sec:det}

In order to perform the measurement to the required precision, a dedicated detector is necessary. We describe here a possible setup to measure the following observables:
\begin{itemize}
\item direction and momentum of the incident muon;
\item directions of the outgoing electron and muon.
\end{itemize}

The CERN muon beam M2, used at 150 GeV, has the characteristics needed for such a measurement.
The beam intensity provides the required event yield.
Its time structure allows to tag the incident muon while keeping low 
 the background related to incoming particles ({\it e.g.} electrons).
The electron contamination is very small.
The beam provides both positive and negative muons, which we plan to use.

The target consists of atomic electrons.
To reach the required statistics, it must contain an adequate 
amount of material to give a sufficient number of electron scattering centres.
The target has to be made of a low-{\it Z} material to minimize the impact of multiple scattering and the background due to bremsstrahlung and pair production processes.

\begin{figure*}[htp]
\begin{center}
\includegraphics[scale=0.55]{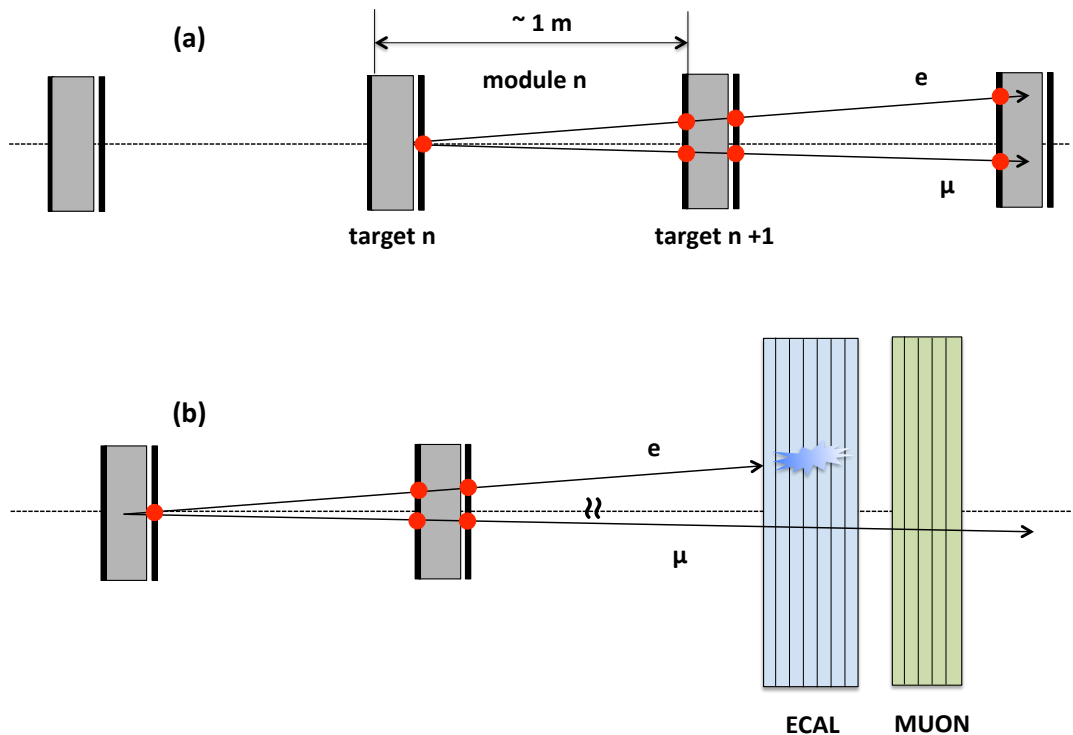}
\caption{Scheme of a possible detector layout. (a) The detector is a modular system. Each module consists of a low-$Z$ target (3~cm of Be or C) 
and two silicon tracking stations located at a distance of one meter. 
(b) To perform the $\mu/e$ discrimination in the case of small scattering angles (both $\theta_{\mu}$ and $\theta_{e}$ below 5 mrad) the detector is equipped with an electromagnetic calorimeter and a muon detector.}
\label{Be-detector}
\end{center}
\end{figure*}

A promising idea, presently under study, is to use 20 identical modules, 
each consisting of a 3~cm thick layer of Be (or C) coupled to 2 Si stations 
located at a relative distance of one meter from each other 
and spaced by intermediate air gaps. 
Fig.~\ref{Be-detector} shows the basic layout.

The arrangement provides both a distributed target with low-{\it Z} and the tracking system. 
As downstream particle identifiers we plan to use a calorimeter for the electrons and a muon system for the muons (a filter plus active planes). 
This particle identifier system is required to solve the muon-electron ambiguity for electron scattering angles around
 (2--3)~mrad 
($cf$. Fig.~\ref{giovanni}).

Preliminary studies of such an apparatus, performed by using GEANT4,
indicate that a tracking angular resolution for the outgoing particles of~$\sim 0.02$~mrad could be 
reached using nowadays available silicon strip detectors.

The detector acceptance covers the region of the signal, with the electron
emitted at extremely forward angles and high energies, as well as the
normalization region, where the electron has much lower energy (around 1 GeV)
and an emission angle of some tens of mrad.

The boosted kinematics of the collision allows the detector to cover almost 100\% of the acceptance, 
and all the scattering angles in the laboratory system to be accessed by a single detector element.

The incoming muons have to be tagged and their direction and momentum
 precisely measured.
To this purpose, a detector similar to those used by COMPASS \cite{compass} or
 NA62 \cite{gtrk_NA62} can be employed.
%
%
%
%
\section{Considerations on systematic uncertainties}
\label{sec:unc}

Significant contributions of the hadronic vacuum polarization to the $\mu e \to \mu e$ differential cross section are essentially restricted to electron scattering angles below 10 mrad, corresponding to electron energies above 10 GeV.
The net effect of these contributions is to increase the cross section by a few
 per mille: a precise determination of $a_{\mu}^{\rm{HLO}}$ requires not only
high statistics, but also a high systematic accuracy, as the final goal of the
experiment is equivalent to a determination of the differential cross section
with $\sim$10 ppm systematic uncertainty at the peak of the integrand function ($cf$. Fig.~\ref{xfunction}).

Such an accuracy can be achieved if the efficiency is kept highly uniform over the entire $q^2$ range, including the normalization region, and over all the detector components.
This motivates the choice of a purely angular measurement: an acceptance of tens of mrad can be covered with a single sensor of modern silicon detectors, positioned at a distance of about one meter from the target.
It has to be stressed that particle identification (electromagnetic
calorimeter and muon filter) is necessary to solve the electron-muon ambiguity
in the region below 5 mrad.
The wrong assignment probability can be measured with the data by using the rate of muon-muon and electron-electron events.

Another requirement for reaching very high accuracy is to measure all
the relevant contributions to systematic uncertainties from the data themselves. 
An important effect, which distinguishes the normalization from the signal
region, is multiple scattering, as the electron energy in the normalization region is
as low as 1 GeV.
Multiple scattering breaks the muon-electron two-body angular correlation, moving events out of the kinematic line in the 2D plot of Fig.~\ref{giovanni}.
In addition, multiple scattering in general causes acoplanarity, while two-body events are planar, within the resolution. 
These facts allow effects to be modelled and measured using data.
 An additional handle on multiple scattering could be the inclusion of a thin
layer in the apparatus, made of the same material as the main target modules. 
This possibility will be studied in detail with simulation. 

The challenge of the proposed measurement is the feasibility of achieving a systematic uncertainty at the level of 10 ppm.
This is the key point from the experimental side. 
In order to demonstrate that such a precision can be realistic, 
a very detailed optimization of the experimental apparatus is necessary.
Tests with beams (electrons and muons), and with one or two modules of the detector, 
will be necessary and a crucial tool to understand if and to what extent the systematic uncertainties can be kept under control.
They will provide a proof-of-concept of the proposed method.
From the first preliminary studies, we are confident that such a challenge can be succesfully addressed.


%
From the theoretical point of view, the control of the systematic
uncertainties requires the development of high-precision Monte Carlo tools, including
all the relevant QED radiative corrections to reach the needed theoretical precision. 

To this aim, QED radiative corrections at fixed
order (NLO, NNLO), properly matched to leading-logarithmic corrections resummed
to all orders of perturbation theory, are mandatory to achieve the necessary theoretical
accuracy on the relevant differential cross sections.

Tools to calculate Bhabha scattering exist, like for instance the
\texttt{BabaYaga} event generator~\cite{babayaga}, which implement exact
NLO corrections matched with leading-logarithmic resummation, ensuring that
the differential cross section is theoretically under control at the
${\mathcal O}(10^{-4})$ level. The same algorithmic framework can be extended to
$\mu e\to\mu e$ scattering and generalized to include exact diagrammatic NNLO corrections. 

For the $\mu e \to \mu e$ case, NLO QED corrections have been
explored in Ref.~\cite{NLOmue} and can be easily reproduced with
modern numerical techniques and tools.
Concerning NNLO corrections, they are not yet available for
$\mu e \to \mu e$ scattering. Nevertheless, the full two-loop result is known for Bhabha
scattering (see Ref.~\cite{Actis:2010gg} and references therein);
we expect that at least some sub-sets of these corrections can be
used ({\it e.g.} two loop corrections which do not connect $e$ and
$\mu$ lines) and the remaining part ({\it e.g.} two-loop box
corrections, which connect the two lines and have two different mass scales)
can be studied and eventually calculated with modern techniques.
Matching NNLO corrections to resummation of higher orders
will shift the theoretical error from contributions of order
$\alpha^2 L$ to order $\alpha^3 L^2$ (where $L$ is a typical collinear
logarithm, {\it i.e.} $L\equiv \log(s/m_e^2)\simeq 14$ for the process under
consideration). From exploratory simulations in the setup of the present proposal in the
case of Bhabha scattering, we estimate that, while ${\mathcal O} (\alpha^2 L)$
contributions on the $t$ distribution are at the level of few
$10^{-4}$, the ${\mathcal O} (\alpha^3 L^2)$ ones are roughly
in the range of $10^{-5}$, therefore reaching the necessary theoretical goal. 

Two comments are in order: first, the above estimate concerns the
Bhabha process and hence can be considered 
as an upper limit of the impact of radiative corrections to the
$\mu e \to \mu e$ process, being the leading collinear logarithm for $\mu$ radiation
($L\simeq 4.5$) smaller than for $e$. Second, when using
the ratio of the cross sections in the
signal and normalization regions, we expect that the theoretical
uncertainty will not deteriorate, due to partial cancellation of common radiative
corrections. 

Work is in progress to extend the available Monte Carlo tools to $\mu e\to\mu e$
scattering and to quantify the achievable accuracy in the computation of
the ratio of signal and normalization cross sections, by means of dedicated and realistic simulations.

\section{Conclusions}
\label{sec:concl}
We presented a novel approach to determine the running of $\alpha$ in the space-like region and $a^{\rm{HLO}}_{\mu}$,
the leading hadronic contribution to the muon $g$-2, by scattering high-energy muons on atomic electrons of a low-$Z$ target through the process $\mu e \to \mu e$.
The experiment is primarily based on a precise measurement of the scattering angles of the two outgoing particles as the $q^2$ of the muon-electron interaction can be directly determined by the electron (or muon) scattering angle.

An advantage of the muon beam is the possibility of employing a modular apparatus, with the target subdivided in subsequent layers.
A low-{\it Z} solid target is preferred in order to provide the required event rate, limiting at the same time the effect of multiple scattering as well as of other types of muon interactions (pair production, bremsstrahlung and nuclear interactions).

The normalization of the cross section is provided by the very same $\mu e \to \mu e$ process in the low-$q^2$ region, where the effect of the hadronic corrections on $\alpha(t)$ is negligible.
Such a simple and robust technique has the potential to keep systematic effects under control,
aiming to reach a systematic uncertainty of the same order as the statistical one.
For this purpose a preliminary detector layout has been described.
By considering a beam of 150~GeV muons with an average intensity of $\sim 1.3\times 10^7$ muon/s, currently available at the CERN North Area, a statistical uncertainty of $\sim 0.3\%$ can be achieved on $a^{\rm{HLO}}_{\mu}$ in two years of data taking.

A test performed using a single detector module, exploiting the muon beam facility, 
could provide a validation of the proposed method.

\begin{acknowledgements}
We like to thank \mbox{Carlo Broggini}, Lau Gatignon, Fred Jegerlehner, Marina Marinkovic and Thomas Teubner for fruitful discussions, 
and \mbox{Fedor Ignatov} for fruitful discussions and help in the simulation.
This work was supported in part by the Italian Ministry of University and Research under the PRIN project 2010YJ2NYW. \\
M.P.\ acknowledges partial support by FP10 ITN Elusives (H2020-MSCA-ITN-2015-674896) and Invisibles-Plus (H2020-MSCA-RISE-2015-690575).
\end{acknowledgements}

%
%

\end{document}